\documentclass[]{aa}
\usepackage{psfig}
\begin{document}

\thesaurus{06(08(08.02.2, 08.06.3, 08.09.2 HV UMa))}

\title{HV~Ursae Majoris, a new contact binary with early-type components
\thanks{Based on the data obtained at the David Dunlap Observatory,
University of Toronto}}

\author{B. Cs\'ak\inst{1}\thanks{Guest Observer, Sierra Nevada 
Observatory} \and L.L. Kiss\inst{2} \and J. Vink\'o\inst{1,}\inst{3}
\and E.J. Alfaro\inst{4}}

\institute{Department of Optics \& Quantum Electronics, University of Szeged,
POB 406, H-6701 Szeged, Hungary
\and Department of Experimental Physics and Astronomical Observatory,
University of Szeged,
Szeged, D\'om t\'er 9., H-6720 Hungary
\and
Research Group on Laser Physics of the Hungarian Academy of Sciences,
Szeged \and
Instituto de Astrof\'\i sica de Andaluc\'\i a, CSIC, P.O. Box 3004,
E-18080, Spain}

\titlerunning{HV Ursae Majoris, a new contact binary}
\authorrunning{Cs\'ak et al.}
\offprints{l.kiss@physx.u-szeged.hu}
\date{received 15 December 1999, accepted 28 January 2000}

\maketitle
 
\begin{abstract}
We present the first $UBV$ and $uvby$ photometric observations for
the short period variable star HV~Ursae~Majoris classified as a
field RRc variable. The observed differences between the consecutive minima
and the lack of colour variations disagree with the RRc-classification
and suggest the possible binary nature of HV~UMa. In order to
reveal the real physical status of this star,
we took medium resolution ($\lambda/\Delta \lambda \approx
11000$) spectra in the red spectral region centered at 6600 \AA.
Spectra obtained around the assumed quadratures clearly
showed the presence of the secondary component.

An improved ephemeris calculated using our and Hipparcos
epoch photometry is Hel. JD$_{\rm min}=2451346.743\pm0.001$,
P$=0\fd7107523(3)$.
A radial velocity curve was determined by modelling the cores of H$\alpha$
profiles with two Gaussian components.
This approximative approach gave a spectroscopic
mass ratio of q$_{\rm sp}$=0.19$\pm$0.03. A modified Lucy model
containing a temperature excess of the secondary was fitted to
the V light curve. The obtained set of physical parameters together
with the parallax measurement indicate that this binary lies
far from the galactic plane, and the primary component is
an evolved object, probably a subgiant or giant star.
The large temperature excess of the secondary may suggest
a poor thermal contact between the components due to a relatively
recent formation of this contact system.

 
\end{abstract}

\section{Introduction}
The first note on the possible light variability
of HV~Ursae~Majoris (= HD~103576 = HIP~58157,
$\langle V \rangle=8.69$, $\Delta V=0.28$, $P=0\fd355385$,
$\pi_{\rm Hipp}=3.12\pm1.23$ mas, $d_{\rm Hipp}=320^{+210}_{-90}$ pc) was
published by Penston (1973) who gave an 'uncertain'
mark to the range of V-magnitude (``var? V=8.60--8.83'').
The periodic nature of the light variation was discovered by 
the Hipparcos satellite (ESA 1997)
and the star was classified as an RRc variable. There is
a note in ESA (1997) about the possibility of a double period but no
firm conclusion was drawn. ESA (1997) gives a spectral type
A3, while Slettebak \& Stock (1959) published A7.

We started a long-term observational project of Str\"omgren photometry and
spectroscopy of the newly discovered bright Hipparcos variables. The
first results have already appeared in Kiss et al. (1999a, b). Since the
period, spectral type and light curve do not exclude the possibility of
wrong classification, accurate determination of the fundamental physical
parameters is highly desirable.

The main aim of this paper is to present the first $UBV$ and $uvby$
photometry for HV~UMa. Also, our radial velocity measurements are
the first time-resolved spectroscopic observations of this
star to date. The paper is organised as follows: the observations
are described in Sect.\ 2, Sect.\ 3 deals with data analysis
and the obtained physical parameters, while a discussion
of the results is given in Sect.\ 4. A final list of conclusions
is presented in Sect.\ 5.

\section{Observations}

\subsection{Photometry}

The Str\"omgren $uvby$ photometric observations
were carried out on 9 nights
in June, 1999, using the
0.9 m telescope at Sierra Nevada Observatory (Spain) equipped with a
six-channel (uvby+$\beta$)
spectrograph photometer (Nielsen 1983).
Earlier, $UBV$ measurements were obtained on one single night
in March, 1999, using the 0.4 m Cassegrain-type telescope of
Szeged Observatory equipped with a single-channel Optec SSP-5A
photometer. These observations covered only 5 hours and
revealed a 0.2 mag variation between two consecutive minima.
We carried out differential photometry with respect to HD~103150
($V$=8.45, $B-V$=0.54, $b-y$=0.335, $m_{\rm 1}$=0.149, $c_{\rm 1}$=0.381 mag).
The overall accuracy of the standard transformation
is about $\pm0.01$ mag for $V$, $b-y$
and $m_1$ and $\pm0.02$ mag for $c_1$.
The light and colour curves were phased using the
corrected ephemeris (see below) and are plotted in Fig.\ 1.

%Table 1.
\begin{table}
\caption{The journal of observations}
\begin{center}
\begin{tabular} {ll}
\hline
Julian Date & type\\
\hline
2451263    & $UBV$\\
2451309    & spectr.\\
2451310    & spectr.\\
2451340    & $uvby$\\
2451341    & $uvby$\\
2451342    & $uvby$\\
2451343    & $uvby$\\
2451345    & $uvby$\\
2451346    & $uvby$\\
2451347    & $uvby$\\
2451349    & $uvby$\\
2451350    & $uvby$\\
\hline
\end{tabular}
\end{center}
\end{table}

%Fig. 1.
\begin{figure}
\begin{center}
\leavevmode
\psfig{figure=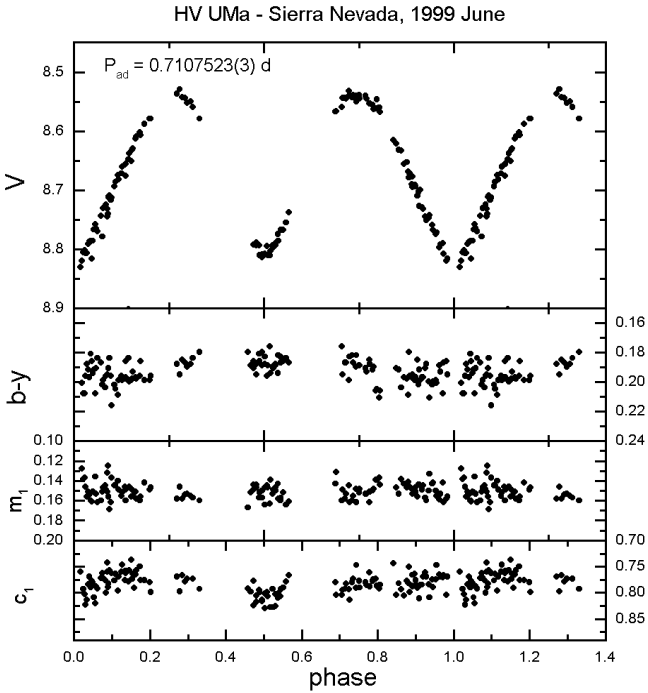,width=\linewidth}
\caption{The light and colour curves of HV~UMa phased with the
adopted ephemeris (see text)}
\end{center}
\label{fig1}
\end{figure}

\subsection{Spectroscopy}

%Fig. 2.
\begin{figure}
\begin{center}
\leavevmode
\psfig{figure=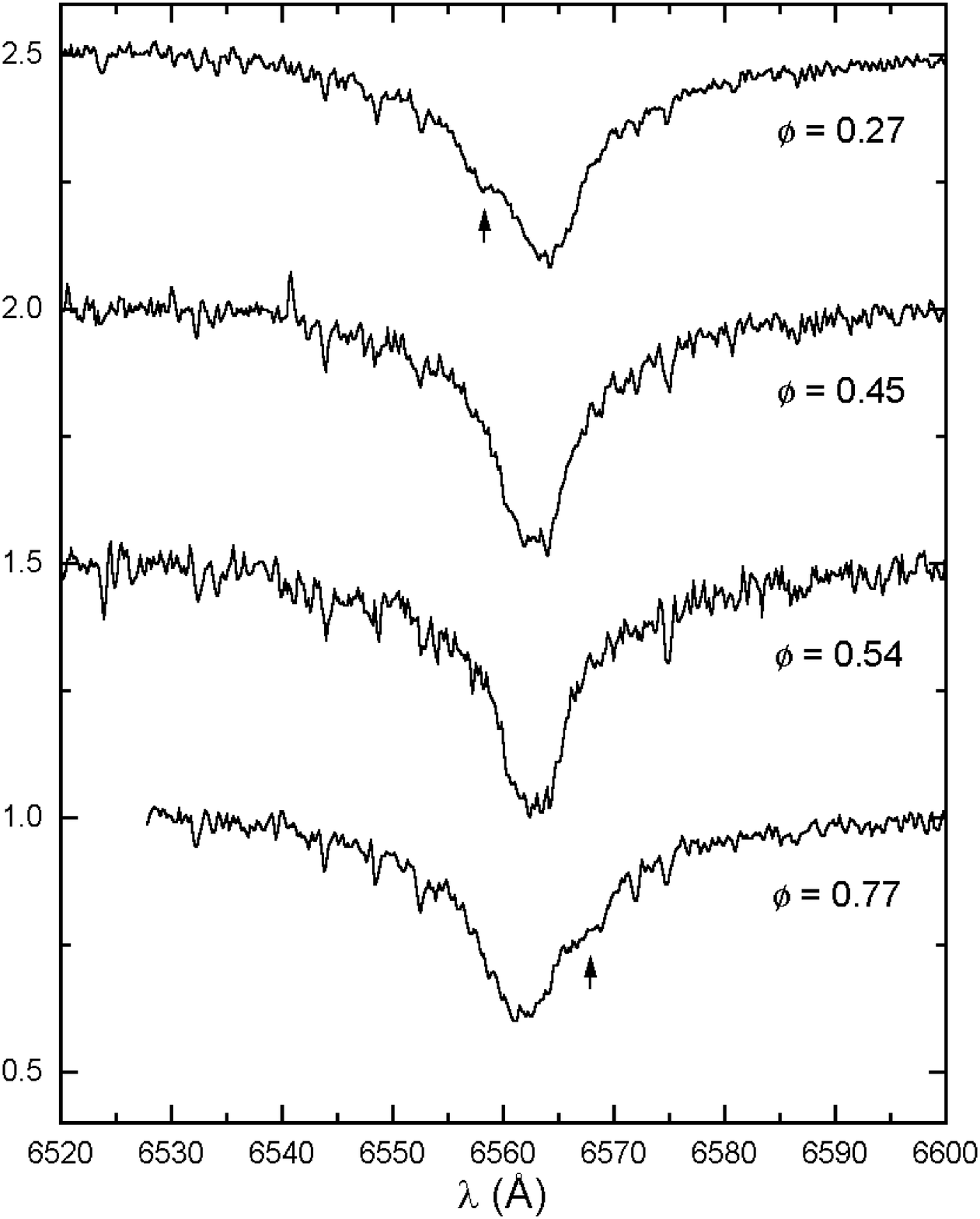,width=\linewidth}
\caption{Sample spectra around the quadratures and minima.
The presence of the secondary component is obvious.}
\end{center}
\label{fig2}
\end{figure}

The spectroscopic observations were carried out at David Dunlap Observatory
with the Cassegrain spectrograph attached to the 74" telescope on two
nights in May, 1999. The detector and the spectrograph setup were the same
as used by Vink\'o et al. (1998). The resolving power
 ($\lambda / \Delta \lambda$) was 11,000 and the signal-to-noise
ratio reached 30--50, depending on the weather conditions.
The spectra were centered on 6600 \AA\ and
reduced with standard IRAF tasks, including bias removal,
flat-fielding, cosmic ray elimination, aperture extraction (with the task
$doslit$) and wavelength calibration. For the latter, two FeAr spectral
lamp exposures were used, which were obtained before and
after every three stellar exposures. The sequence of observations
FeAr-var-var-var-FeAr was chosen because of the short period of
HV~UMa. Careful linear interpolation between the two comparison
spectra was applied in order to take into account the sub-pixel
shifts of the three stellar spectra caused by the movement of
the telescope. We chose an exposure time of 10 minutes, which corresponds
to 0.01 in binary orbital phase, avoiding phase smearing of the radial
velocity curve. The spectra were normalized to the continuum by fitting
a cubic spline, omitting the region of H$\alpha$.

Besides a few telluric features, only the H$\alpha$ line
could be detected with acceptable S/N ratio in our
200 \AA-wide spectra. At the phases of maximum light
the $H\alpha$ line exhibited significant broadening and
an excess bump appeared on the wings alternating between
the blue and the red side (see Fig.\ 2). It can be interpreted
most easily as the effect of a close companion star,
therefore HV UMa is most probably a spectroscopic binary.
Fig.\ 2 shows a few sample spectra with the calculated
orbital phase indicated on the right side of every spectrum.

\section{Physical parameters}

\subsection{Epoch and period}

The Hipparcos data suggested that the light variation
of HV~UMa can be described with a single period of
$0\fd355385$ (ESA 1997). We observed only
one moment of minimum (Hel. JD = 2451346.388), but the consecutive
minimum appeared to be slightly fainter, therefore, we adopted a
doubled Hipparcos period as a first approach ($0\fd71077$) and shifted
the observed time of minimum with $0\fd355385$ to obtain the final
epoch Hel. JD$_{\rm min}=2451346.743\pm0.001$.

The next step was to refine the period. This was done by
phasing Hipparcos epoch photometry with the newly determined
epoch and the doubled Hipparcos-period. The resulting phase diagram
showed a shift of $\Delta \phi \approx 0.1$ (=$0\fd071$).
That shift was eliminated by recalculating the period until
correct phase diagrams for both our and Hipparcos data (Fig.\ 3) were
reached.
The resultant period is $P=0\fd7107523(3)$. The fact that
earlier Hipparcos data agree very well with our data suggests a
quite stable period of HV~UMa.

%Fig. 3.
\begin{figure}
\begin{center}
\leavevmode
\psfig{figure=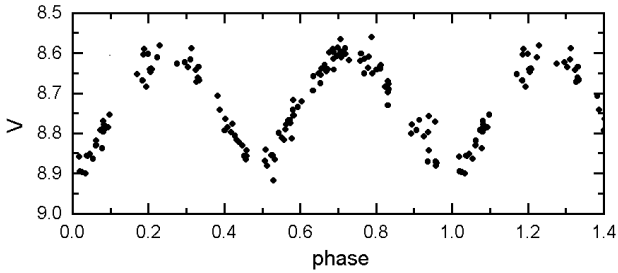,width=\linewidth}
\caption{Hipparcos epoch photometry data phased with
the finally adopted ephemeris (see text)}
\end{center}
\label{fig3}
\end{figure}

\subsection{Classification}

The shape of the light curve, i.e. the continuous light
variation and the very deep secondary minimum (almost
as deep as the primary one), the absence of significant
colour variation, the appearance of the secondary line in the spectra
at the quadrature phases all suggest that HV~UMa is probably
an eclipsing contact binary. This is confirmed
by its low mass ratio and a consistent model of the
light curve (see below).

Comparing the light curve phased with the final epoch
and period (Fig.\ 1) with the line profiles
observed at phases of maximum light (Fig.\ 2) it is visible
that the secondary line appears on the {\it blue} side
at the $\phi = 0.27$ quadrature phase that follows the
deeper minimum, while this bump is redshifted at
$\phi = 0.77$. This indicates that the smaller companion star
approaches us after the primary minimum, therefore that
minimum is due to an occultation eclipse. The weak
point of this analysis is that the light curve is
not very well covered around the minima either by
the Hipparcos data or our observations. The
data obtained at Sierra Nevada have better inner precision
(less scatter) than those provided by Hipparcos,
and these data indicate that the
minimum at $\phi = 0.0$ is slightly deeper. Therefore,
we adopted this eclipse as primary minimum, but this needs
further confirmation. If the deeper minimmum is really
due to an occultation eclipse then HV~UMa is a so-called
W-type contact binary.

Contact binaries can contain early (O--B) or
late (G--K) spectral type stars. The latter group is
referred to as the W~UMa stars, while the former is known
as early-type contact systems. The colours of HV~UMa
indicate early F spectral type, therefore HV~UMa is
an ``intermediate'' type contact binary between the
W~UMa stars and the OB-type contact systems. The
surface temperature and the line profile of the HV~UMa
system makes it similar to the known contact systems
UZ~Leo and CV~Cyg (Vink\'o et al., 1996).

\subsection{Radial velocities and spectroscopic mass ratio}

Since the secondary component is only partly resolved, the
radial velocities must be determined by modelling the individual
line profiles in order to avoid blending effects (e.g. systematic
decrease of the velocity amplitude). For this purpose
we chose those spectra that were obtained around the quadratures.
These show the presence of the secondary most clearly. One spectrum
around light minimum was also modelled to test the applied method.

Because the H$\alpha$ profile is strongly affected by the
Stark broadening and shows wide non-gaussian wings, we
normalized the profiles to the surrounding continuum,
and selected
the lower part of the profiles below the 0.9 intensity value.
We fitted two individual Gaussian profiles to the line cores adjusting
the amplitudes, FWHM values and line core positions. The initial
values of these parameters were estimated from two spectra very close to
the quadratures ($\phi$=0.25 and 0.77). 
The FWHM converged very quickly to the final values, being 7.6 \AA\ and
4.0 \AA\ for the primary and secondary components, respectively.
Line depths changed slightly from spectrum to spectrum, as the
contributions are phase-dependent, resulting in 0.29--0.30 for the primary
and 0.07--0.10 for the secondary (note, that these values mean
line depths below 0.9 normalized intensity). The fitted
line core positions resulted in the radial velocity variations for
both components. Sample spectra with the fitted profile are shown
in Fig.\ 4, while the radial velocities are presented in Table\ 2.
The estimated accuracy of the individual velocities is about 5 km~s$^{-1}$ for the
primary, and 10 km~s$^{-1}$ for the secondary, which is mainly
determined by the resolution of the line core in wavelength.
The velocity amplitudes resulting from this method
are 47$\pm$1.5 km~s$^{-1}$ and 254$\pm$10
km~s$^{-1}$, where the uncertainties
are due to the random errors caused
by the observational scatter. The corresponding mass-ratio is
$q_{\rm sp}=m_{\rm sec}/m_{\rm pri}=0.185\pm0.01$.

However, as was also pointed out by the referee, this kind of
velocity determination may contain a large amount of systematic
error, mainly due to the assumed Gaussian shape of the individual
line profiles. The intrinsic H$\alpha$ profiles of the components
of HV~UMa are probably quite different from Gaussian, therefore
this approach can be considered as only the first approximation
for extracting the radial velocities from the H$\alpha$ profiles.
The major part of the systematic error is governed by the shape of the 
wing of the primary component's model profile on the side where the
secondary star appears (blueward at $\phi = 0.25$ and redward
at $\phi = 0.75$ phases). It is well visible in Fig.\ 4 that the
position of the secondary line is shifted toward larger velocities with
respect to the position of the ``hump'' on the observed profile,
due to the increasing contribution of the primary line toward
the main minimum of the combined line. If the primary line was steeper on 
the side where the secondary line exists, overlapping the secondary
by a smaller amount, then the
secondary line would be less shifted, thus, its position would be
closer to the local hump on the observed profile, resulting in 
a smaller radial velocity of the secondary. On the other hand,
a shallower secondary profile would give us systematically
higher velocities due to the same reason.

In order to estimate
the amount of this kind of systematic error, we simply determined
the positions of the two local minima (the main minimum and the 
secondary's hump) on the profiles observed around quadratures 
(four spectra around $\phi = 0.25$ and two around $\phi = 0.75$)
when the presence of the hump appeared to be most prominent.
This was done interactively, by eye, plotting the line profiles
on the computer screen, which again introduced some subjectivity
into the procedure, but it is stressed that this is done only
for estimating the {\it errors} of the velocities and not 
for obtaining their actual values. Of course, the velocities
of the secondary measured in this way were systematically smaller
than those obtained by the Gaussian fitting. The velocities
of the primary were almost the same, as could be expected.
The total amplitude turned out to be $K' = 280$ km~s$^{-1}$, while 
the mass ratio changed to $q' = 0.22$. Comparing these values
with the results of the Gaussian fitting, we conclude that
the errors of the radial velocity amplitude and the spectroscopic
mass ratio (both random and systematic) are approximately $\pm 23$ 
km~s$^{-1}$ and $\pm 0.03$, respectively. The finally adopted
parameters determined spectroscopically, together with their
errors are collected in Table\ 3. It is important to note
that the mass ratio can be refined by modelling the light curve
(Sect. 3.4), but the total velocity amplitude is tied only 
to the spectroscopic data, thus, its uncertainty will directly
appear in the absolute parameters of the system.

%Table 2.
\begin{table}
\caption{The observed heliocentric radial velocities obtained
by the Gaussian fit.
The velocity resolution is about 5 km~s$^{-1}$.}
\begin{center}
\begin{tabular} {llrr}
\hline
Hel. JD    & $\phi$   & $V_{\rm rad}(prim.)$ & $V_{\rm rad}(sec.)$ \\
2400000+  &          &   [km/s]             &    [km/s]              \\
\hline
51309.6206 & 0.77     &   $-$47            & 251\\
51309.6396 & 0.80     &   $-$51            & 242\\
51309.6471 & 0.81     &   $-$42            & 246\\
51310.6411 & 0.21     &   41             & $-$252\\
51310.6485 & 0.22     &   45             & $-$243\\
51310.6562 & 0.23     &   50             & $-$243\\
51310.6643 & 0.24     &   54             & $-$238\\
51310.6719 & 0.25     &   50             & $-$243\\
51310.6794 & 0.26     &   50             & $-$243\\
51310.6893 & 0.27     &   45             & $-$238\\
51310.6964 & 0.28     &   41             & $-$252\\
51310.7037 & 0.29     &   45             & $-$247\\
51310.7436 & 0.35     &   41             & $-$233\\
51310.8423 & 0.49     &    9             &  9\\
\hline
\end{tabular}
\end{center}
\end{table}

%Fig. 4.
\begin{figure}
\begin{center}
\leavevmode
\psfig{figure=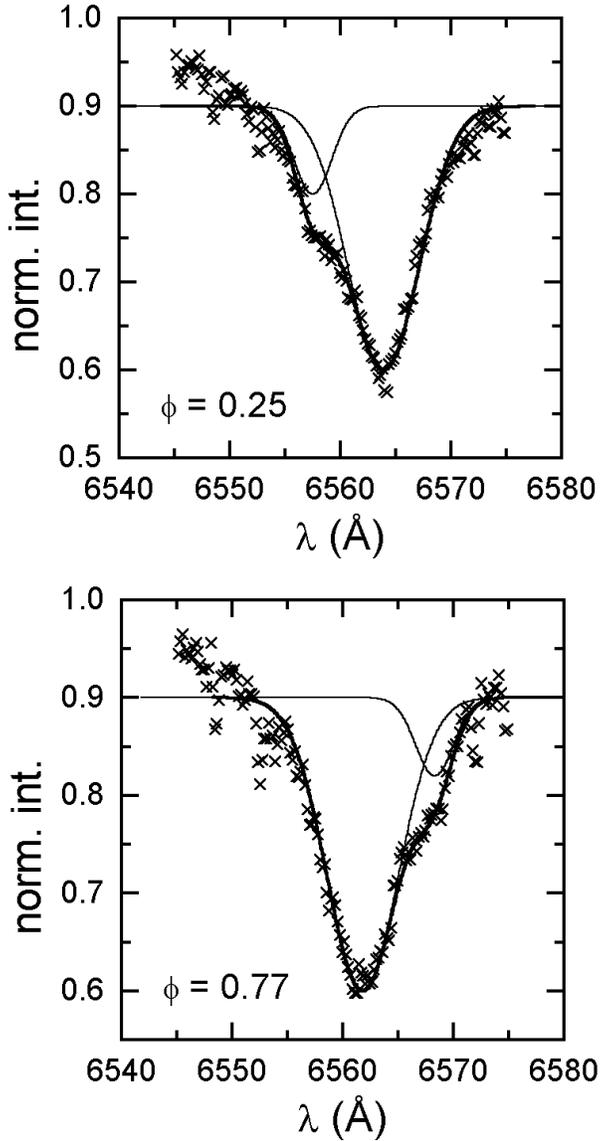,width=\linewidth}
\caption{The observed (crosses) and fitted (solid lines) Gaussian
line profiles at two quadrature phases.}
\end{center}
\label{fig4}
\end{figure}

\subsection{Light curve modelling}

The V-light curve was synthesized with the computer
code {\it BINSYN} described briefly in Vink\'o et al. (1996).
This code is based
on the usual Roche-model characterized by the
geometric parameters $q_{\rm ph}$ (photometric mass-ratio),
$F$ (fill-out) and $i$ (orbital inclination).
The relative depth
of the eclipses were modelled introducing the relative
temperature excess of the secondary
$X = (T_{\rm sec} - T_{\rm pri})/ T_{\rm pri}$ (hot-secondary model).
Because the primary minimum turned out to be due to occultation,
the phases were shifted by 0.5 assigning $\phi = 0.0$
to the transit eclipse (built-in default in {\it BINSYN}).

%Fig. 5.
\begin{figure}
\begin{center}
\leavevmode
\psfig{figure=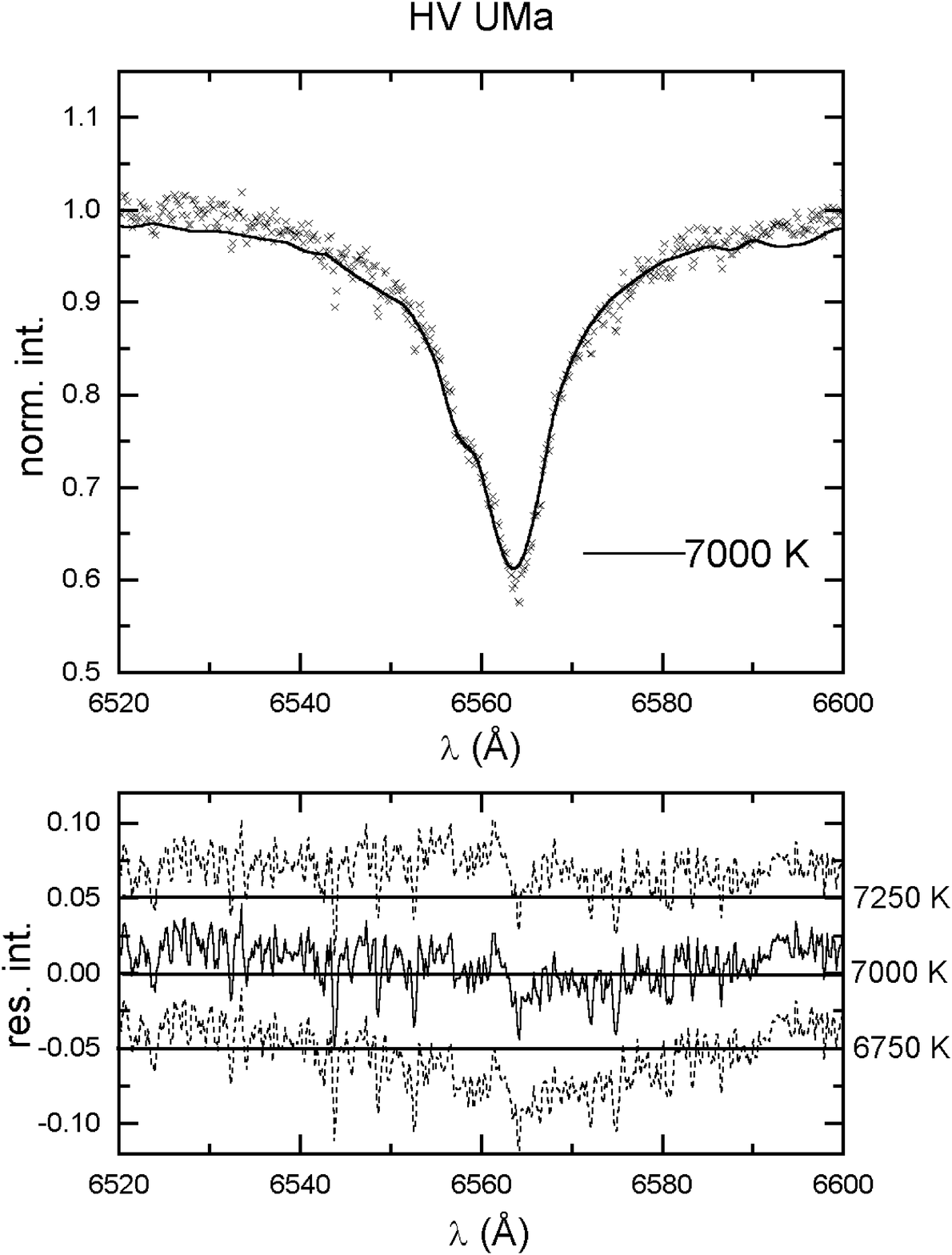,width=\linewidth}
\caption{The observed (crosses) and calculated H$\alpha$ profiles at
$\phi$=0.25. The spectrum was calculated by an ATLAS9 code adopting
$T_{\rm eff}=7000$ K, log~$g$=4.0 and full radial velocity amplitude
of 300 km~s$^{-1}$. The bottom panel shows the residual intensities
for the adopted fit compared with two other models (6750 K and 7250 K).
The overall agreement is the best for $T_{\rm eff}=7000$ K. The sharp
absorption features are atmospheric telluric lines.}
\end{center}
\label{fig5}
\end{figure}

First, the effective temperature of the primary component
was estimated based on synthetic colour
grids by Kurucz (1993) and the observed mean Str\"omgren colour
indices ($\langle b-y \rangle = 0.19$ mag, $ \langle m_1 \rangle
= 0.15$ mag,
$\langle c_1 \rangle = 0.77$ mag), resulting in $T_{\rm eff} = 7300 \pm 200$
K and $\log g = 4.0 \pm 0.3$ dex (assuming $E(B-V) = 0$
and solar chemical abundance). The interstellar reddening
in the direction and distance of HV UMa is expected to be small,
because this variable lies far from the galactic plane.
TU UMa, an RR Lyr variable lying 17$^\circ$ SE from HV UMa also has a
negligible colour excess (Liu \& Janes, 1989).

Other parameters necessary for modelling the binary star
were as follows.
A linear limb-darkening law with coefficient
$u = 0.61$ was adopted from tables of Al-Naimiy (1977).
The gravity darkening exponent and the bolometric
albedo were chosen at their usual values for
radiative atmospheres: $\beta = 0.25$ and $A = 1.0$.
All these parameters were kept fixed during the
solution for the best light-curve model.

The light-curve fitting was computed using a controlled
random search method, the so-called Price algorithm
(Barone et al., 1990, Vink\'o et al., 1996).
The optimized parameters were $q_{\rm ph}$, $F$, $i$ and
$X$. The best solution was searched for in the following
parameter domains: $0.05 < q < 0.5$, $1.01 < F < 1.99$,
$50 < i < 90$ and $<-0.2 < X < +0.2$. The fit quickly
converged to low inclination and low mass ratio values
that were expected from the shallow eclipses
($\Delta V \approx 0.3$ mag) and
the small spectroscopic mass ratio ($q_{\rm sp} = 0.19$).
Also, it turned out that there are strong correlations
between the optimized parameters.
Due to this correlation, the physical parameters
determined from the light curve fitting cannot
be considered as a unique solution: certain
parameter combinations describe the light curve
almost equally well. In these cases the combination
of photometric and spectroscopic information is
very important: one can use the spectroscopic
data to determine a consistent set of physical
parameters that gives the best model fitting all
the available data.

In order to combine the photometric and spectroscopic
information and find a consistent description of the system,
we modelled the observed H$\alpha$ line profiles using
the parameters from the light curve fitting.
The model line profiles were computed by
convolving an intrinsic H$\alpha$ profile of a
non-rotating star with the Doppler-broadening profile
of the contact binary. The broadening profiles were
calculated with the WUMA4 code (Rucinski, 1973).
For the determination of the intrinsic H$\alpha$ profile
we used Kurucz's ATLAS9 code modified by John Lester.
This approach, however, has some limitations, because
the H$\alpha$ line is strongly affected by NLTE-mechanisms,
therefore the ATLAS9-model profile will be somewhat
different from the real intrinsic profile, especially
near the line core. Thus, only a crude comparison
between the modelled and observed line profiles was
possible, neglecting the differences in the line core.

%Fig. 6.
\begin{figure}
\begin{center}
\leavevmode
\psfig{figure=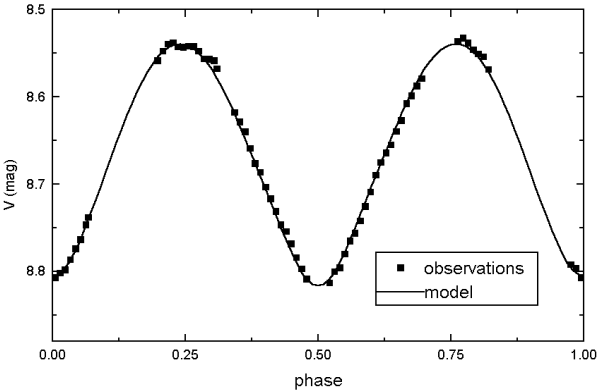,width=\linewidth}
\caption{Fitted light curve of HV~UMa. Note a 0.5 shift in phase
to get a transit eclipse at $\phi$=0.}
\end{center}
\label{fig6}
\end{figure}

However, it turned out that the originally adopted
effective temperature $T_{\rm eff} = 7300$ K was too high.
With this temperature, the Stark-wings of the H$\alpha$
line are so strong that they overwhelm the rotational
broadening, producing much wider line profiles than
observed. Thus, we reduced the value of the effective
temperature until satisfactory agreement was found
between the observations and the broadened model profiles.
The resulting fit is plotted in Fig.\ 5 together with the
observed line at $\phi = 0.25$.
The residuals of three model profiles are also shown.
It can be seen that
at $T_{\rm eff} = 6750$ K the wings can be fitted quite
well, but the computed line core is too shallow.
On the other hand, if $T_{\rm eff} = 7250$ K is used,
the computed line core agrees better, but the wings
are wider. Thus, $T_{\rm eff} = 7000$ K was accepted
as a compromise. The Str\"omgren-colours of the
Kurucz-grid for this temperature are almost the
same as for the originally adopted 7300 K, therefore
this temperature is still consistent with the
photometric colours.

The light curve modelling was recomputed with
$T_{\rm eff} = 7000$ K. The optimized parameters changed
only slightly, resulted in a slightly smaller mass ratio
and a slightly higher temperature excess. Table\ 3 shows
the final set of physical parameters, while the fitted
light curve is visible in Fig.\ 6. The distribution
of the random points around the $\chi^2$-minimum in
the parameter space is plotted in Fig.\ 7. The structure
of the sub-spaces in this diagram indicates the
correlation between the different parameters.

%Fig. 7.
\begin{figure}
\begin{center}
\leavevmode
\psfig{figure=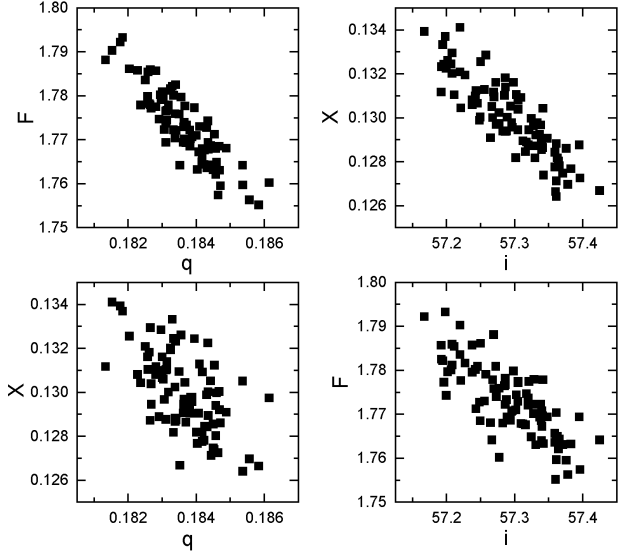,width=\linewidth}
\caption{Distribution of random points around the minimum of $\chi^2$.}
\end{center}
\label{fig7}
\end{figure}

Another light curve model was computed in order to test
the effect of the gravity darkening and reflection
parameters that were originally fixed as if the
atmosphere was radiative. The model with their
``convective'' values $\beta = 0.08$ and $A = 0.5$
resulted in an even larger temperature excess than
in the radiative case. Because the temperature excess
of the secondary is only a ``correction'' parameter
in the light curve solution and it may not mean
real temperature difference, it would be difficult
to explain physically a very large value of the temperature
excess that still does not cause significant colour
variation. Thus, we adopted the model
with radiative atmospheric parameters as our final
solution, and this model is listed in Table\ 3.
Note that the errors
of the fitted parameters (3rd column) are difficult to estimate,
because of the parameter correlation. We monitored the behaviour
of the $\chi^2$ function during the optimization and assigned
uncertainties to each parameter according to the spread of
the random points for which $\chi^2~<~0.5$. This criterion
defines those solutions when the fitted curve runs well within
the error bar at each measured normal point, giving a feasible
fit to all observations. The parameter correlation means that
the uncertainties are also not independent of each other:
e.g. slightly decreasing $q$ forces increasing $F$ or $X$ 
(see Fig.7). The uncertainties of the calculated parameters
(3rd panel in Table 3) were estimated assuming a $\pm23$ km~s$^{-1}$
error in the radial velocity amplitude $K$.

Because of the correlation between the optimized
parameters, it is very important to check whether
the radial velocities calculated from the model
match the observed velocities. This comparison
is plotted in Fig.\ 8 where an almost perfect
agreement can be seen. The low mass ratio results
in the distortion of the sinusoidal velocity curves,
which is a well-known effect in close binaries.

%Table 3
\begin{table}
\caption{Physical parameters of HV~UMa.}
\begin{center}
\begin{tabular} {lll}
\hline
Fitted parameters          & Value  & $\sigma$ \\
\hline
{\it spectroscopy} & &\\
\hline
$K$~(km~s$^{-1})$ & 300 &   23\\
$q_{sp}$         & 0.19 &  0.03\\
$T_{eff}$~(K)  & 7000  &  200\\
\hline
{\it photometry} & &\\
\hline
$q_{ph}$    &   0.184  & 0.05\\
$F$       &   1.77   & 0.15\\
 $i$      &   57.3   &  0.4\\
$X$       &   0.13   & 0.03\\
\hline
Calculated parameters      &  Value & $\sigma$ \\
\hline
$a$ (10$^6$ km)      & 3.48   &  0.25 \\
$M_1 (M_\odot)$      & 2.8    &  0.6 \\
$M_2 (M_\odot)$      & 0.5    &  0.17 \\
$R_1 (R_\odot)$      & 2.62   &  0.25 \\
$R_2 (R_\odot)$      & 1.18   &  0.16 \\
$\rho_{pri}$~(g~cm$^{-3}$)   & 0.2    &  0.05 \\
\hline
\end{tabular}
\end{center}
\end{table}

%Fig. 8.
\begin{figure}
\begin{center}
\leavevmode
\psfig{figure=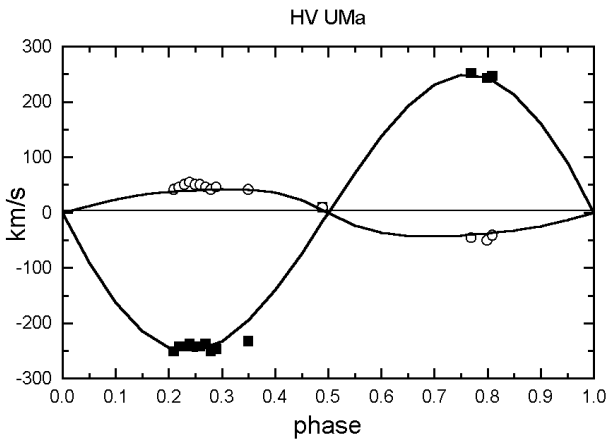,width=\linewidth}
\caption{The radial velocity curves of the primary and secondary components.
Symbols correspond to the directly measured values, while solid lines
denote the calculated ones from the photometric model.}
\end{center}
\label{fig8}
\end{figure}

\section{Discussion}

The spectroscopic detection of the secondary component
and the success of modelling the light- and velocity
curves as well as the H$\alpha$ line profile supports
our conclusion that HV~UMa is not a short-period
RR Lyr variable, but an eclipsing binary system.
The physical parameters listed in Table\ 3 give
a deep contact configuration of this binary, explaining
naturally the lack of significant colour change
during the light variation cycle, which would be
peculiar in a pulsating variable. Since the primary
minimum is due to occultation, HV~UMa is a so-called
W-type contact binary (systems exhibiting transit
eclipses as primary minima are called A-type).

It is well known that contact binaries can be
formed either from hot, early-type stars or
cool, late-type stars, the latter representing the
class of W~UMa-type variables (see e.g. Rucinski, 1993;
Figueiredo et al., 1994 for reviews).
The surface temperatures of W~UMa-systems are
usually below 7000 K, while the temperatures
of early-type contact binaries show a much wider
range, between 10,000 and 40,000 K. Thus, HV~UMa
falls into the temperature regime between 7000 and
10,000 K that is relatively rarely occupied by
contact binaries. Systems like HV~UMa may
represent a transition population between early-type
and late-type contact binaries (although even the
existence of such transition is questionable).
These systems
lie on the boundary between radiative and convective
atmospheres, at about late A - early F spectral
types. Detailed studies of such
systems would be important, because the formation
and the structure of early-type systems having radiative
envelopes and late-type contact binaries having
convective envelopes is probably quite different.

The period value $P \approx 0.711$ day
also indicates that HV~UMa is not a typical contact
system. Recent statistical studies based on
the data from the $OGLE$ microlensing survey
(Rucinski, 1998 and references therein) have shown
that the number of contact systems strongly decreases
above $P = 0.7$ day, and there is a well-defined limit
at $P \approx 1.5$ days. Therefore, HV~UMa is a member
of the relatively rare ``longer-period'' contact
binaries, although a few contact systems with
$P > 2$ days definitely exist, at least close to the
galactic bulge. On the other hand, there exists a
period-colour relation of ``normal'' W~UMa-stars
with period $P < 1$ day, which indicates that longer
period systems have bluer colour. On the plot of
the empirical $\log P - (b-y)_0$ relation (Rucinski,
1983) the position of HV UMa, using the mean colour
$\langle b-y \rangle = (b-y)_0 = 0.19$ (Fig.1, assuming zero reddening)
and our revised period $\log P = -0.1483$ (Sect. 3.1),
is close to the upper boundary of the relation,
suggesting an atypical, but not peculiar contact system.

A more recent
$\log P - (B-V)_0$ diagram based on Hipparcos-parallaxes
has been published by Rucinski (1997, see his Fig.2).
The position of HV~UMa on this diagram was calculated 
assuming $E(B-V) = 0$ and $(b-y) / (B-V) = 0.7$, 
resulting in $(B-V)_0 \approx 0.27$. These data show that
the position of HV~UMa is entirely consistent with that of
involving the older $\log P - (b-y)$ relation, being close to
the blue short-period envelope (BSPE, Rucinski, 1997),
but the system is definitely redder than the upper limit
defined by the BSPE, consistently with other contact binaries.
Also, a rough comparison of HV~UMa with other contact binaries
on the $\log P - (V-I)_0$ diagram based on $OGLE$-photometry 
(Rucinski, 1998) strengthens the status of HV~UMa outlined above,
again, being closer to the BSPE than other systems with similar
period,
although the lack of observed $(V-I)$ colour of HV~UMa limits
the reliability of this comparison at present.  
It can be concluded that all available measurements and pieces
of information consistently support the contact binary nature of
HV~UMa.

The new physical parameters collected in Table\ 3, together with
the Hipparcos-parallax ($\pi = 3.12 \pm 1.23$ mas) enable
us to estimate the evolutionary status of HV~UMa, provided it
is indeed a contact binary with those parameters. The coordinates
and the distance based on the Hipparcos-parallax indicate
that HV UMa is a halo object: its distance from the galactic
plane is $z \approx 300$ pc, which means that $[$Fe/H$] = 0$ 
(assumed during the analysis of the line profiles and the colour 
indices) may not be true. Because the spectroscopic observations
presented in this paper have limited spectral range 
($\Delta \lambda = 200$ \AA\ around 6600 \AA), and this region
does not contain significant metallic lines in early-type stars,
the spectroscopic derivation of $[$Fe/H$]$ was not possible.
Thus, because of the lack of further information we assumed
$[$Fe/H$] = 0$, which is not impossible for halo objects, but
$[$Fe/H$] < 0$ is also likely.

\begin{figure*}
\begin{center}
\leavevmode
\psfig{file=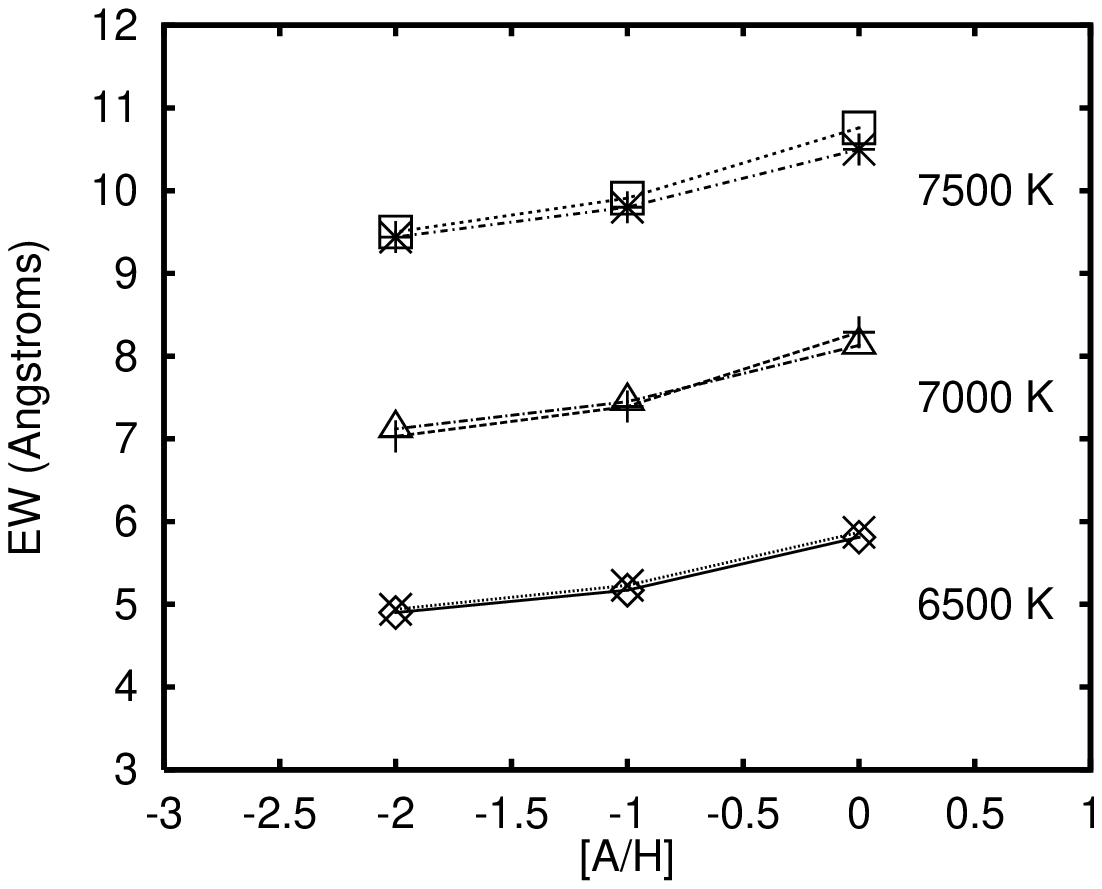,height=5cm,width=7cm}
\psfig{file=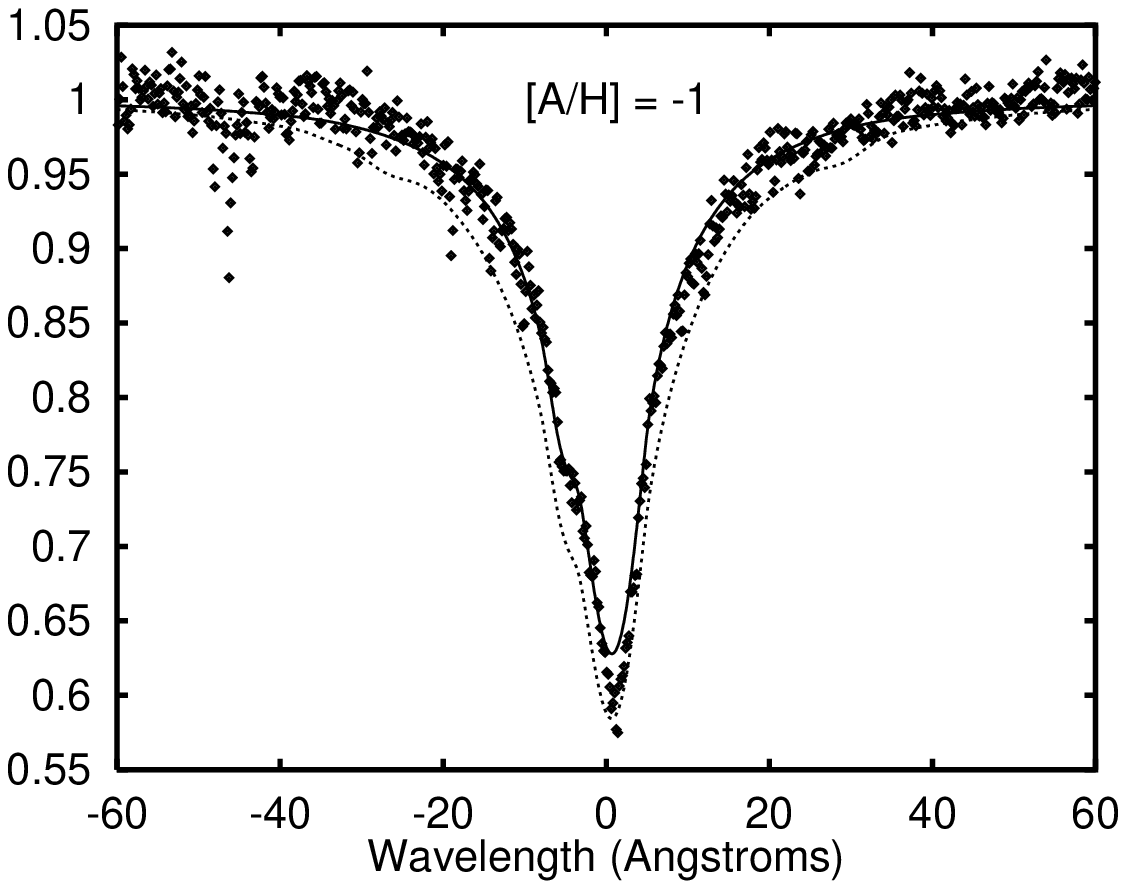,height=5cm,width=7cm}
\end{center}
\caption{Effect of metallicity on the H$\alpha$ line profile. Left panel:
the dependence of the H$\alpha$ equivalent width on the metallicity
for three temperatures (labeled) and two $\log g$ values (4 and 3.5)
for each temperature. Right panel: comparison of the H$\alpha$ of low
metallicity models (for $T_{\rm eff} = 7000$ and $7500$ K)
with the observed profile.}
\end{figure*}

The referee suggested the possibility that
the low metal content of HV~UMa critically affects the
Stark broadening of the hydrogen lines, thus, significantly
influencing the temperature derived from the H$\alpha$ profile
in Sect.\ 3.4. We investigated this effect in detail using
the pre-computed H$\alpha$ profiles by Kurucz (1979) including
Stark broadening. In the left panel of Fig.\ 9 the dependence
of the H$\alpha$ equivalent width on metallicity is plotted.
For each temperature, two model sequences for $\log g = 4.0$
and 3.5 are shown.
It can be seen that the decrease of the metallicity indeed
affects the strength of H$\alpha$, but in this temperature
range the variation of the equivalent width is governed mainly
by the change of the effective temperature. The gravity (pressure)
dependence is very weak. Because the equivalent
width of the broad H$\alpha$ line strongly depends on the
strength of the Stark wings, it is expected that the wings
of the H$\alpha$ profile presented in Fig.\ 5 (corresponding
to $[$A/H$] = 0$) are not affected very largely by the possible
lower metallicity of HV~UMa, thus, the derived temperature
$T_{\rm eff} = 7000$ K is only slightly dependent on metal content.
This is illustrated in the right panel of Fig.\ 9, where two model
profiles corresponding to two different temperatures 
($T_{\rm eff} = 7000$ and $7500$ K) and $[$A/H$] = -1.0$ i.e.
significantly lower metallicity than assumed in the previous section
are presented together with the observed line profile at quadrature.
It can be seen that the $T_{\rm eff} = 7500$ K model still gives
a broader line profile than observed, while the $T_{\rm eff} = 7000$ K
model results in a much better agreement in the wings, very similar
to the case of solar metallicity presented in Fig.\ 5. Note, however,
that the lower metal content causes a less deep line core of H$\alpha$,
thus, the problem of fitting the whole H$\alpha$ line is 
exaggerated when the effect of metallicity is taken
into account. Nevertheless, it is concluded that the 
$T_{\rm eff} = 7000 \pm 200$ km~s$^{-1}$ temperature derived from
the wings of H$\alpha$ probably does not contain a significant
systematic error due to the unknown metallicity of HV~UMa.

As was mentioned in Sect.\ 3.4, the colour excess of HV~UMa
is $E(B-V) \approx 0$. This is supported by the effective temperature
derived spectroscopically (discussed above) and 
photometrically (from observed and tabulated Str\"omgren indices), 
because both methods resulted in a consistent value. The 
negligible reddening is also in agreement with the statement that
HV~UMa belongs to the halo population.

At first glance, the absolute geometric parameters collected in Table\ 3
would indicate that the HV~UMa system consists of main-sequence
components: both stars have $\log g = 4.0$ and the mass and radius values
of the primary are also similar to those of a main sequence star (the
secondary is oversized in relation to its mass, typical of contact systems).
However, the surface temperatures and luminosities indicate that HV~UMa is
probably an evolved object. First, the combined absolute magnitude of the
system based on parallax measurement and $E(B-V) = 0$ results in 
$M_{\rm V} = 1.0 \pm 0.8$ mag, where the large error is due to the uncertainty
of the Hipparcos-parallax. Using tabulated bolometric corrections, the
total luminosity of the system is $L_{\rm T} = 30 \pm 20 L_\odot$.
Second, the luminosities of the components
are $L_{\rm i} = 4 \pi R_{\rm i}^2 \sigma T_{\rm eff}^4 = 15.4 L_\odot$
and $2.9
L_\odot$ for the primary and secondary, respectively, giving
$L_{\rm T} = 18.3 L_\odot$ for the combined luminosity, which is
within the error of the distance-based total luminosity estimated above.
However, both of these luminosities are much less than the expected
luminosity $L \approx 60 \pm 10 L_\odot$ of a main sequence star with
$M \approx 3 M_\odot$ (Lang, 1991). Moreover, this kind of main sequence
star would have $T_{\rm eff} = 9500$ K, much higher than the surface
temperature of HV~UMa. Therefore, the primary component of HV~UMa
is too cool and too faint for its mass if it is assumed to be
a main sequence object.

The agreement with a class III giant star having 
$L \approx 30 \pm 10 L_\odot$ for $M \approx 3 M_\odot$ is much better.
The temperature of 
such giant star is $T_{\rm eff} \approx 7400 \pm 300$ K which is not
very far from the surface temperature of HV~UMa. Taking into
account the energy transfer between the components in the
contact binary (assuming that the total luminosity of the system is due to
the energy production of only the more massive primary component), 
the corrected effective temperature of
the primary component is $T_{\rm 1,corr} = 7300 \pm 100$ K. The
radius and the surface gravity of this giant star, 
$R = 2.9 R_\odot$ and $\log g \approx 3.8$, also agrees well
with the derived parameters of HV~UMa. Therefore, the
comparison of empirical and theoretical values of the physical 
parameters suggests that the primary component of HV~UMa is
an evolved object, probably a IV-III class subgiant, or giant star.
Because W~UMa stars are generally accepted to belong to the
old disk population (e.g. Rucinski, 1993, 1998), it is reasonable
that a long-period contact system, containing a more massive
primary than most of other W~UMa systems, is significantly evolved
from the main sequence. Therefore, the evolved status of HV~UMa
qualitatively agrees with the age of other contact binaries.

It is interesting to compare the direct empirical absolute magnitude 
of HV~UMa derived above ($M_V = 1.0$ mag) with the prediction of
the period-colour-luminosity relation of W~UMa stars calibrated
by Rucinski \& Duerbeck (1997) as $M_{\rm V} = 0.10 + 3.08 (B-V)_0 - 4.42 \log P$.
Using the same estimated $(B-V)_0$ index as above, the
predicted absolute magnitude for HV~UMa becomes 
$M_{\rm V} = 1.59 \pm 0.35$ mag, which agrees with
the empirical value within the errors. Note, that the deviation
of some of the calibrating W~UMa stars in the sample of Rucinski
\& Duerbeck (1997) from the value predicted by this relation is as
large as $0.5 - 0.7$ mag (see Fig.\ 4 in Rucinski \& Duerbeck 1997),
therefore the difference between
the observed and the predicted absolute magnitude of HV~UMa 
does not make this system discrepant with respect to other contact
binaries. On the other hand, it is a bit surprising that the 
relation that is mainly based on main sequence objects 
gave such a good prediction for the more evolved HV~UMa system. 
This agreement is probably limited to the particular range on
the HR-diagram close to the position of HV~UMa, and may not
hold on for more evolved systems with $P > 1$ day. 
Very few known contact systems exist above the $P = 1$ day period
value, as recently discussed by Rucinski (1998), this lack of
systems also gives a natural limit for the applicability of this
relation for longer periods.

The separation of the components in the HV~UMa system and
the evolved physical state of the primary may suggest that
this contact system formed during a case B mass transfer.
This may also give a reasonable explanation for the poor
thermal contact $\Delta T = 900$ K between the
components. Model computations of the formation of
contact binaries via evolution induced mass transfer
from the more massive component (Sarna \& Fedorova, 1989) 
predict large amount of temperature excess
($\Delta T \approx 2000 - 3000$ K) at the moment of reaching
the contact configuration. The result that the eclipse
depths of HV~UMa can be modelled with only such high
temperature excess may indicate that this contact
system formed only recently and did not have enough time
to reach better thermal contact. Note, that the temperature
excess in W~UMa-type contact binaries is usually considered
unphysical, because the lack of the colour index variation
suggests very good thermal contact for late-type stars.
The physically consistent model of the eclipse depths of 
W~UMa-stars contains large starspots on the surface of one
or both components (e.g. Hendry et al., 1992). However,
in the case of HV~UMa with $T_{\rm eff} >= 7000$ K,
the presence of such starspots is less likely, thus, the
eclipse depths of this system may indeed mean a 900 K
temperature difference between the secondary and the primary.

\section{Conclusions}

The new results presented in this paper can be summarized
as follows.

\noindent 1. We reported the first $uvby$ photometric and medium
resolution spectroscopic observations of the short-period
variable star HV~Ursae~Majoris. Contrary to the RRc classification
given by ESA (1997), the star turned out to be a new
contact binary with early-type components.

\noindent 2. An improved ephemeris was determined using our
and Hipparcos epoch photometric data:
Hel.JD$_{min}$=2451346.743+0.7107523(3)$\cdot$E. There is
no indication of changing period over almost 10 years.

\noindent 3. A radial velocity curve was measured directly from
the H$\alpha$ profiles by two-component Gaussian fitting of the line core
regions in spectra recorded around the quadratures. We
calculated a spectroscopic mass ratio of 0.19$\pm$0.03.

\noindent 4. The effective temperature and the surface gravity were
determined using the mean Str\"omgren indices, synthetic
colours of Kurucz (1993) and theoretical (ATLAS9)
line profiles giving
$T_{eff}=7000\pm200$ K and log~$g=4.0\pm0.3$.
The light curve modelling resulted in a complete set of physical parameters.

\noindent 5. The physical parameters of HV UMa together with the
parallax measurement indicate that this binary is situated
far from the galactic plane, and the primary component
is an evolved object, probably a subgiant or giant star.
The large temperature excess of the secondary may be
indicative of a poor thermal contact between the
components due to a relatively recent formation of this
contact system via case B mass transfer.

\begin{acknowledgements}
This research was supported by MTA-CSIC Joint Project No.15/1998,
Hungarian OTKA Grants \#F022249, \#T022259, \#T032258,
Pro Renovanda Cultura
Hungariae Foundation Grant DT 1999 \'apr./36. and
Szeged Observatory Foundation.
LLK wishes to express his thanks to the staff of the DDO for
granting the necessary observing time. Also, LLK and
BCS acknowledge the helpful assistance by \'E. Bar\'at and B. Gere
during the observations. Fruitful discussions with
K. Szatm\'ary are also gratefully acknowledged.
Thanks are due to the referee, Prof. S. Rucinski,
whose criticism and many suggestions led to significant
improvement of the paper.
The NASA ADS Abstract
Service was used to access data and references.
\end{acknowledgements}


\begin{thebibliography}{}

\bibitem[1977]{alnaim77}
    Al-Naimiy H.M. 1977, Ap\&SS 66, 281

\bibitem[1990]{barone}
    Barone F., Milano L., Russo G. 1990, in: Active Close
    Binaries (ed. C.Ibanoglu), Kluwer Acad. Publ., p. 161.

\bibitem[1997]{esa97}
    ESA 1997, The Hipparcos and Tycho Catalogues, ESA SP-1200

\bibitem[1994]{figu94}
    Figueiredo J., De Greve J.P., Hilditch R.W. 1994, A\&A 283, 144

\bibitem[1992]{hendry92}
    Hendry P.D., Mochnacki S.W., Collier Cameron A. 1992, ApJ 399, 246

\bibitem[1999]{kiss99a}
    Kiss L.L., Cs\'ak B., Thomson J.R., Szatm\'ary K. 1999a,
    IBVS No. 4660

\bibitem[1999]{kiss99b}
    Kiss L.L., Cs\'ak B., Thomson J.R., Vink\'o J. 1999b,
    A\&A 345, 149

\bibitem[1979]{kur79}
    Kurucz R.L. 1979, ApJ S. 40, 1

\bibitem[1993]{kurucz}
    Kurucz R.L. 1993, ATLAS9 Stellar Atmosphere Programs and 2 km/s
    Model Grids, CR-ROM No.13

\bibitem[1991]{lang91}
    Lang K.L. 1991, Astrophysical Data: Planets and Stars, Springer-Verlag

\bibitem[1989]{liu89}
    Liu T., Janes K.A. 1989, ApJS 69, 593

\bibitem[1983]{niels83}
    Nielsen R. F. 1983, Institute Theoric. Astrophysics Oslo Report No. 59,
    ed. O. Hauge, p 141

\bibitem[1973]{penston}
    Penston M.J. 1973, MNRAS 164, 133

\bibitem[1973]{ruc73}
    Rucinski S.M. 1973, Acta Astron 23, 79

\bibitem[1983]{ruc83}
    Rucinski S.M. 1983, A\&A 127, 84

\bibitem[1993]{ruc93}
    Rucinski S.M. 1993, in: The Realm of Interacting Binary Stars
    eds. J Sahade et al., Kluwer, p.111.

\bibitem[1997]{ruc97a}
    Rucinski S.M. 1997, AJ 113, 1112

\bibitem[1997]{ruc97b}
    Rucinski S.M., Duerbeck H.W. 1997, PASP 109, 1340

\bibitem[1998]{ruc98}
    Rucinski S.M. 1998, AJ 115, 1135

\bibitem[1989]{sarna89}
    Sarna M.J., Fedorova A.V. 1989, A\&A 208, 111

\bibitem[1959]{slett}
    Slettebak A., Stock J. 1959, Astron. Abh. Hamburg Sternwarte 5, 105

\bibitem[1996]{vinko96}
    Vink\'o J., Heged\"us T., Hendry P.D. 1996, MNRAS 280, 489

\bibitem[1998]{vinko98}
    Vink\'o J., Evans N.R., Kiss L.L., Szabados L. 1998, MNRAS, 296, 824

\end{thebibliography}
\end{document}